\documentstyle{article}
\begin{document}
\def\np#1#2#3{Nucl. Phys. {\bf B#1} (#2) #3}
\def\pl#1#2#3{Phys. Lett. {\bf #1B} (#2) #3}
\def\prl#1#2#3{Phys. Rev. Lett.{\bf #1} (#2) #3}
\def\physrev#1#2#3{Phys. Rev. {\bf D#1} (#2) #3}
\def\ap#1#2#3{Ann. Phys. {\bf #1} (#2) #3}
\def\prep#1#2#3{Phys. Rep. {\bf #1} (#2) #3}
\def\rmp#1#2#3{Rev. Mod. Phys. {\bf #1}}
\def\cmp#1#2#3{Comm. Math. Phys. {\bf #1} (#2) #3}
\def\mpl#1#2#3{Mod. Phys. Lett. {\bf #1} (#2) #3}
\def\tilde{\widetilde}
\def \eq#1{\begin{equation}#1\end{equation}}
\def \tb#1{\left(\begin{array}#1\end{array}\right)}
\def \abs#1{\left|#1\right|}
\def \bra#1{\left\langle #1\right\vert}
\def \ket#1{\left\vert #1\right\rangle}
\def \exp{\mbox{exp}}
\def \tp{\mbox{tp\,}}
\def \ignoruj#1{}
\def \eqn#1#2{\eq{#2\label{#1}}}
\def \ha{{1\over 2}}
\quad\vspace{10mm}
\begin{flushright}
     hep-th/9703218\\
     RU-97-17\\ HEP-UK-0004
\end{flushright}
\vspace{10mm}

\begin{center}
 {\LARGE \bf Heterotic Strings
from Matrices}\footnote{\normalsize February 1997}\\
 \vspace{15mm}
 {\Large \bf Tom Banks}\\
 \vspace{3mm}

{\it Department of Physics and Astronomy}\\
{\it Rutgers University}\\
{\it Piscataway, NJ 08855-0849}\\
{\tt banks@physics.rutgers.edu}\\
\vspace{5mm}
 {\Large \bf Lubo\v s Motl}\\
 \vspace{3mm}

{\it Faculty of Mathematics and Physics}\\
{\it Charles University}\\
{\it Prague, Czech Republic}\\
{\tt motl@karlin.mff.cuni.cz}

\vspace{12mm}
{\Large \bf Abstract}
\vspace{3mm}

We propose a nonperturbative definition\\
of heterotic string theory
on arbitrary multidimensional tori.
\end{center}

\newpage  
\tableofcontents
\section{Introduction}

The matrix model of uncompactified M theory \cite{dhn}-\cite{bfss}
has been generalized to
arbitrary toroidal compactifications of type IIA and IIB string theory.
These models can be viewed as particular large $M$ limits of the
original matrix model, in the sense that
they may be viewed as the dynamics of a restricted class
of large $M$ matrices, with the original matrix model Lagrangian.

A separate line of reasoning has led to a description of the
Ho\v rava-Witten domain wall in terms of matrix quantum
mechanics \cite{orbi}. Here,
extra degrees of freedom have to be added to the original matrix model.
As we will review below, if these new variables, which tranform in the
vector representation of the gauge group, are not added, then the model
does not live in an eleven dimensional spacetime, but only on its
boundary..  Although it is, by
construction, a unitary quantum mechanics, it probably does not recover
ten dimensional Lorentz invariance in the large $M$ limit.  Its nominal
massless particle content is the ten dimensional $N = 1$ supergravity
(SUGRA) multiplet, which is anomalous.

With the proper number of vector variables added, the theory does have
an eleven dimensional interpretation.  It is possible to speak of states
far from the domain wall and to show that they behave exactly like the
model of \cite{bfss}.  Our purpose in the present paper is to compactify
this model on spaces of the general form $S^1 / Z_2 \times T^d$.  We
begin by reviewing the argument for the single domain wall quantum
mechanics, and generalize it to an $S^1 / Z_2$ compactification.  The
infinite momentum frame Hamiltonian for this system is practically
identical to the static gauge $O(M)$ Super Yang Mills (SYM)
Hamiltonian for $M$ heterotic D strings
in Type I string theory.  They differ only in the boundary conditions
imposed on the fermions which transform in the vector of $O(M)$.  These
fermions are required for $O(M)$ anomaly cancellation in both models,
but the local anomaly does not fix their boundary conditions.
Along the moduli space of the $O(M)$ theory, the model exactly
reproduces the string field theory Fock space of
heterotic string theory.  The inclusion of both Ramond and Neveu Schwarz
boundary conditions for the matter fermions, and the GSO projection, are
simple consequences of the $O(M)$ gauge invariance of the model.

Generalizing to higher dimensions,
we find that the heterotic matrix model on
$S^1 / Z_2 \times T^d$ is represented by a $U(M)$ gauge theory on
$S^1 \times T^d / Z_2$ .  On the orbifold  circles, the gauge
invariance reduces to $O(M)$.  We are able to construct both the
heterotic and open string sectors of the model, which dominate in
different limits of the space of compactifications.

In the conclusions, we discuss the question of whether the heterotic
models which we have constructed are continuously connected to the
original uncompactified eleven dimensional matrix model.  The answer to
this question leads to rather surprising conclusions, which inspire us
to propose a conjecture about the way in which the matrix model solves
the cosmological constant problem.  It also suggests that string vacua
with different numbers of supersymmetries are really states of {\it
different} underlying theories.  They can only be continuously connected
in limiting situations where the degrees of freedom which differentiate
them decouple.

\section{Heterotic Matrix Models in Ten and Eleven Dimensions}

In \cite{evashamit} an $O(M)$ gauged supersymmetric
 matrix model for a single Ho\v rava-Witten domain
wall embedded in eleven dimensions was proposed.  It was based on an
extrapolation of the quantum mechanics describing $D0$ branes near an
orientifold plane in Type IA string theory \cite{df}.  The model was
presented as an orbifold of the original \cite{bfss} matrix model in
\cite{motl}.
 In the Type IA context
it is natural to add degrees of freedom transforming in the vector of
$O(M)$ and corresponding to the existence
of $D8$ branes and the $08$ strings connecting them to the $D0$ branes.
Since $D8$ branes are movable in Type IA theory, there are consistent
theories both with and without these extra degrees of freedom.  That is,
we can consistently give them masses, which represent the distances
between the $D8$ branes and the orientifold.
However,
as first pointed out by \cite{df}, unless the number of $D8$ branes sitting
near the orientifold is exactly $8$, the $D0$ branes feel a linear
potential which either attracts them to or repels them from the
orientifold. This is the expression in the quantum mechanical approximation,
of the
linearly varying dilaton first found by Polchinski and Witten \cite{jopo}.
This system was studied further in \cite{rk} and
\cite{bss}. In the latter work the supersymmetry and gauge structure of
model were clarified, and the linear potential was shown to correspond to
the fact that the ``supersymmetric ground state''
of the model along classical flat directions representing excursions
away from the orientifold was not gauge invariant.

{}From this discussion it is clear that the only way to obtain a model
with an eleven dimensional interpretation is to add sixteen massless
real fermions
transforming in the vector of $O(M)$, which is the model proposed in
\cite{evashamit}.   In this case, $D0$ branes can move freely away from the
wall, and far away from it the theory reduces to the $U([{M\over 2}])$
model of \cite{bfss} \footnote{Actually there
is a highly nontrivial question
which must be answered in order to prove that the effects of the
wall are localized.
In \cite{bss} it was shown that supersymmetry allowed an arbitary
metric for the coordinate representing excursions away from the wall.
In finite orders of perturbation theory the metric falls off with
distance but, as in the discussion of the graviton scattering amplitude
in \cite{bfss}, one might worry that at large $M$ these could sum up to
something with different distance dependence.  In \cite{bfss} a
nonrenormalization theorem was conjectured to protect the relevant term
in the effective action.  This cannot be the case here.}.

Our task now is to construct a model representing two Ho\v rava-Witten
end of the world $9$-branes separated by an interval
of ten dimensional space.
As in \cite{motl} we can approach this task by attempting to mod out the $1+
1$ dimensional field theory \cite{bfss}, \cite{taylor}, \cite{motlb},
\cite{bs}, \cite{dvv} which describes M
theory compactified on a circle.  Following the logic of \cite{motl}, this
leads to an $O(M)$ gauge theory. The $9$-branes are stretched around
the longitudinal direction of the infinite momentum frame (IMF)
and the $2-9$ hyperplane of the transverse
space.  $X^1$ is the differential operator
$${R_1 \over i}{\partial \over \partial\sigma} - A_1$$
where $\sigma$ is in $[0,2\pi ]$, and $A_1$ is
an $O(M)$ vector potential.  The other $X^i$ transform in the ${\bf
{M(M+1) \over 2}}$ of $O(M)$.  There are two kinds of fermion multiplet.
$\theta$ is an ${\bf 8_c}$ of the spacetime $SO(8)$, a symmetric tensor
of $O(M)$ and is the superpartner of $X^i$ under the eight dynamical
and eight kinematical SUSYs which survive the projection.  $\lambda$ is
in the adjoint of $O(M)$, the ${\bf 8_s}$ of $SO(8)$, and is the
superpartner of the gauge potential.  We will call it the gaugino.

As pointed out in \cite{rk} and \cite{bss}, this model is anomalous.
One must add $32$ Majorana-Weyl fermions $\chi$ in the ${\bf M}$ of $O(M)$.
For sufficiently large $M$, this is the only fermion content which can
cancel the anomaly.   The continuous $SO(M)$ anomaly does not fix the
boundary conditions of the $\chi$ fields. There are various consistency
conditions which help to fix them, but in part we must make a choice
which reflects the physics of the situation which we are trying to
model.

The first condition follows from the fact that our gauge group is $O(M)$
rather than $SO(M)$.  That is, it should consist of the subgroup of
$U(M)$ which survives the orbifold projection.  The additional $Z_2$
acts only on the $\chi$ fields, by reflection.  As a consequence, the
general principles of gauge theory  tell us that each $\chi$ field
might appear with either periodic or antiperiodic boundary conditions,
corresponding to a choice of $O(M)$ bundle.  We must also make a
projection by the discrete transformation which reflects all the
$\chi$'s.
 What is left undetermined by these
principles is choice of relative boundary
conditions among the $32$ $\chi$'s.

The Lagrangian for the $\chi$ fields is
\eqn{chilag}{\chi (\partial_t
+ 2\pi R_1\partial_{\sigma} - i A_0 - i A_1) \chi
.}  In the large $R_1$ limit, the volume of the space on which the gauge
theory is compactified is small, and its coupling is weak, so we can
treat it by semiclassical methods.  In particular, the Wilson lines
become classical variables.  We will refer to classical values of the
Wilson lines as expectation values of the gauge potential $A_1$.
(We use the term expectation value loosely, for we are dealing with a
quantum system in finite volume.  What we mean
is that these ``expectation values'' are the slow variables in
a system which is being treated by the Born-Oppenheimer approximation.)
An excitation of the system at some position in the direction tranverse
to the walls is represented by a wave function of
$n \times n$ block matrices in which $A_1$
has an expectation value breaking $O(n)$ to $U(1) \times U([n/2])$.
In the presence of a generic expectation value, in $A_0 = 0$
gauge, the $\chi$ fields will not have any zero frequency modes.
The exceptional positions where zero frequency modes exist are $A_1 = 0$
(for periodic fermions) and $A_1 = \pi R_1$ (for antiperiodic fermions).
These define the positions of the end of the world $9$-branes, which we
call the walls.  When $R_1 \gg l_{11}$, all of the finite wavelength
modes of all of the fields have very high frequencies and can be
integrated out.
In this limit, an excitation
far removed from the walls has precisely the degrees of freedom of a
$U([{n\over2}])$ gauge quantum mechanics.  The entire content of the
theory far from
the walls is $U([{M\over 2}])$ gauge quantum mechanics.  It has no
excitations carrying the quantum numbers created by the $\chi$ fields,
and according to the conjecture of \cite{bfss} it reduces to eleven
dimensional M theory in the large $M$ limit.  This reduction assumes
that there is no longe range interaction between the walls and the rest
of the system.

In order to fulfill this latter condition it must be true that at $A_1 =
0$, and in the large $R_1$ limit, the field theory reproduces
the $O(M)$ quantum mechanics described at the beginning of this section
(and a similar condition near the other boundary).
We should find $16$ $\chi$ zero modes near each wall.  {\it Thus,
the theory must contain a sector in which
 the $32$ $1+1$ dimensional $\chi$ fields are grouped
in groups of $16$ with opposite periodicity}. Half of the fields will
supply the required zero modes near each of the walls.  Of course, the
question of which fields have periodic and which antiperiodic boundary
conditions is a choice of $O(M)$ gauge.  However, in any gauge
only half of the $\chi$ fields will have zero modes located at any
given wall.   We could of course consider sectors of the
fundamental $O(M)$ gauge theory in which there is a different
correlation between boundary conditions of the $\chi$ fields.
However, these would not have an eleven dimensional interpretation at
large $R_1$.  The different sectors are not mixed by the Hamiltonian so
we may as well ignore them.

To summarize, we propose that M theory compactified on $S^1 / Z_2$ is
described by a
$1+1$ dimensional $O(M)$ gauge theory with $(0,8)$ SUSY.  Apart from the
$(A_{\mu}, \lambda )$ gauge multiplet, it contains
a right moving $X^i, \theta$ supermultiplet in the symmetric tensor
of $O(M)$ and 32 left moving fermions, $\chi$, in the vector.  The
allowed gauge bundles for $\chi$ (which transforms under the discrete
reflection which leaves all other multiplets invariant), are those in
which two groups of $16$ fields have opposite periodicities.
In the next section we will generalize this construction to
compactifications on general tori.

First let us see how heterotic strings emerge from this formalism in the
limit of small $R_1$.  It is obvious that in this limit, the string
tension term in the SYM Lagrangian becomes very small.  Let us rescale
our $X^i$ and time variables so that the quadratic part of the
Lagrangian is independent of $R_1$.  Then, as in \cite{bs},
\cite{motlb}, \cite{dvv},
the commutator term involving the $X^i$ gets a coefficient
$R^{-3}$ so that we are forced onto the moduli space in that limit.  In
this $O(M)$ system, this means that the $X^i$ matrices are diagonal,
and the gauge group is completely broken
to a semidirect product of $Z_2$ (or
$O(1)$) subgroups which reflect the individual components of the
vector representation, and an $S_M$ which permutes the eigenvalues of
the $X^i$.  The moduli space of low
energy
fields\footnote{We use the term moduli space to refer to the space of
low energy fields whose effective theory describes the small $R_1$
limit (or to the target space of this effective theory).
These fields are in a Kosterlitz Thouless phase and do not have
expectation values, but the term moduli space is a convenient shorthand
for this subspace of the full space of degrees of freedom.}
consists of diagonal $X^i$ fields, their superpartners $\theta_a$ (also
diagonal matrices), and the $32$ massless left moving $\chi$ fields.
The gauge bosons and their superpartners $\lambda^{\dot{\alpha}}$
decouple in the small $R_1$ limit.  All of the $\chi$ fields are light
in this limit.

\subsection{Screwing Heterotic Strings}

As first explained in \cite{motlb} and elaborated in
\cite{bs}, and \cite{dvv},
twisted sectors under $S_N$ lead to strings of arbitrary
length\footnote{These observations are mathematically identical to
considerations that arose in the counting of BPS states in black hole
physics \cite{BPS} . }.
The
strings of conventional string theory, carrying continuous values of the
longitudinal momentum, are obtained by taking $N$ to infinity
and concentrating on
cycles whose length is a finite fraction of $N$.
The new feature which arises in the heterotic string is that the
boundary conditions of the $\chi$ fields can be twisted by the discrete
group of reflections.

A string configuration of length $2\pi k$, $X_S^i (s)$, $0 \leq s \leq
2\pi k$, is represented by a diagonal matrix:

 \eqn{screw}{
X^i(\sigma)=\tb{{cccc}
X_S^i(\sigma)&&&\\
&X_S^i(\sigma+2\pi)&&\\
&&\ddots&\\
&&&X_S^i(\sigma+2\pi(N-1))}.}
This satisfies the twisted boundary condition
$X^i(\sigma+2\pi)=E_O^{-1}X^i(\sigma) E_O$ with
\eqn{bcx}{
E_O=\tb{{ccccc}
&&&&\epsilon_k\\
\epsilon_1&&&&\\
&\epsilon_2&&&\\
&&\ddots&&\\
&&&\epsilon_{N-1}&},}
and $\epsilon_i = \pm 1$.
The latter represent the $O(1)^k$ transformations,
which of course do not effect $X^i$ at all.

To describe the possible twisted sectors of the matter fermions we
introduce the matrix $r^a_b = diag (1\ldots 1 ,-1 \ldots -1)$, which
acts on the $32$ valued index of the $\chi$ fields.
The sectors are then defined by
\eqn{chibc}{\chi^a (\sigma + 2\pi ) = r^a_b E_O^{-1} \chi^b
(\sigma )}

As usual, inequivalent sectors
correspond to conjugacy classes of the gauge group.
In this case, the classes can be described by a permutation
with a given set of cycle
lengths, corresponding to a collection of
strings with fixed longitudinal momentum fractions,
and the determinants of the $O(1)^k$ matrices inside each cycle.
In order to understand the various gauge bundles,
it is convenient to write the ``screwing
formulae'' which express the
components of the vectors $\chi^a$ in terms of string
fields $\chi_s^a$ defined on the interval $[0, 2\pi k]$.
The defining boundary conditions are
\eqn{bcchi}{\chi_i^a (\sigma + 2\pi ) =
\epsilon_i r^a_b \chi_{i + 1}^b (\sigma )}
where we choose the gauge in which $\epsilon_{i < k} =1$
and $\epsilon_k = \pm 1$ depending
on the sign of the determinant.  The vector index $i$ is counted modulo $k$.
This condition is solved by
\eqn{soln}{\chi_i^a (\sigma ) = (r^{i - 1})^a_b \chi_S^b
(\sigma + 2\pi (i - 1))}
where $\chi_S$ satisfies
\eqn{hetbc}{\chi_S^a (\sigma + 2\pi k)
= (r^k)^a_b \epsilon_k \chi_S^b (\sigma )}
For $k$ even, this gives the PP and AA sectors
of the heterotic string, according to the
sign of the determinant.
Similarly, for $k$ odd, we obtain the AP and PA sectors.

As usual in a gauge theory,
we must project on gauge invariant states.  It turns out that
there are only two independent kinds of
conditions which must be imposed.  In a sector characterized by a
permutation $S$, one can be chosen to be
the overall multiplication of $\chi$ fields associated with a given
cycle of the permutation (a given string)
by $-1$.
This GSO operator anticommuting with all the 32 $\chi$ fields
is represented by the ${\bf -1}$ matrix from the gauge group $O(N)$.
The other is the projection associated with the cyclic
permutations themselves.
It is easy to verify that under the latter transformation
the $\chi_S$ fields transform as
\eqn{chitransf}{\chi_S^a (\sigma )
\rightarrow r^a_b \chi_S^b (\sigma + 2\pi)}
Here $\sigma \in [0,2\pi k]$ and we are
taking the limit $M\rightarrow \infty$,
$k/M$ fixed.  In this limit the $2\pi$
shift in argument on the righthand side of
(\ref{chitransf}) is negligible,
and we obtain the second GSO projection of the heterotic string.

Thus, $1+1$ dimensional $O(M)$
SYM theory with $(0,8)$ SUSY, a left moving
supermultiplet in the symmetric
tensor representation and 32 right moving fermion
multiplets in the vector (half
with P and half with A boundary conditions)
reduces in the weak coupling, small (dual) circle limit to two copies of
the Ho\v rava-Witten domain wall
quantum mechanics, and in the strong coupling large (dual)
circle limit, to the string field theory of the $E_8 \times E_8$
heterotic string.

\section{Multidimensional Cylinders}

The new feature of heterotic compactification on $S^1 /Z_2 \times T^d$
is that the coordinates in the toroidal dimensions are represented by
covariant derivative operators with respect to new world volume
coordinates.  We will reserve $\sigma$ for the periodic coordinate dual
to the interval $S^1 / Z_2$ and denote the other coordinates by
$\sigma^A$.   Then,
\eqn{covder}{X^A = {2\pi R_A \over i}{\partial
\over \partial \sigma^A} - A_A
(\sigma ); \quad A = 2\ldots k+1.}
Derivative operators are antisymmetric, so in order to implement the
orbifold projection, we have to include the transformation $\sigma^A
\rightarrow  - \sigma^A$, for $A = 2\ldots d + 1$,
 in the definition of the orbifold symmetry.
Thus, the space on which SYM is compactified is $S^1 \times (T^d /
Z_2)$.   There are $2^d$ {\it orbifold circles} in this space, which are
the fixed manifolds of the reflection.
Away from these singular loci, the gauge group
is $U(M)$ but it will be restricted to $O(M)$ at the singularities.
We will argue that there must be
a number of $1 + 1$ dimensional fermions living only on these circles.
When $d = 1$ these orbifold lines can be thought of as the boundaries of
a {\it dual cylinder}.
Note that if we take $d = 1$ and rescale the $\sigma^A$ coordinates
so that their lengths are $1/R_A$ then a long thin cylinder in spacetime
maps into a long thin cylinder on the world volume, and a short fat
cylinder maps into a short fat cylinder.  As we will see, this
geometrical fact is responsible for the emergence of Type $IA$
and heterotic strings in the appropriate limits.

The boundary conditions on the world volume fields are
\eqn{bca}{X^i (\sigma ,\sigma^A ) = \bar X^i (\sigma ,  - \sigma^A ),\quad
A_a (\sigma ,\sigma^A ) = \bar A_a (\sigma ,  - \sigma^A ),}
\eqn{bcc}{A_1 (\sigma ,\sigma^A ) = -\bar A_1 (\sigma ,  - \sigma^A)}
\eqn{bcd}{\theta (\sigma ,\sigma^A ) = \bar\theta(\sigma ,-\sigma^A),\quad
\lambda (\sigma ,\sigma^A ) = -\bar\lambda (\sigma ,-\sigma^A )}
All matrices are hermitian,
so transposition is equivalent to complex conjugation.
The right hand side of the boundary condition \ref{bcc} can also be shifted
by $2\pi R_1$, reflecting the fact that $A_1$ is an angle variable.

Let us concentrate on the cylinder case, $d=1$.
In the limits $R_1 \ll l_{11} \ll R_2$ and
 $R_2 \ll l_{11} \ll R_1$, we will find that
the low energy dynamics is completely
described in terms of the moduli space,
which consists of commuting $X^i$ fields.
In the first of these limits,
low energy fields have no $\sigma^2$ dependence, and
the boundary conditions restrict
the gauge group to be $O(M)$, and force
$X^i$ and $\theta$ to be real symmetric matrices.
Anomaly arguments
then inform us of the existence of $32$
fermions living on the boundary circles.
The model reduces to the $E_8 \times E_8$
heterotic matrix model described in the previous
section, which, in the indicated limit,
was shown to be the free string field theory of
heterotic strings.

\subsection{Type IA Strings}

The alternate limit produces something novel.
Now, low energy fields are restricted
to be functions only of $\sigma^2$.
Let us begin with a description of closed strings.
We will exhibit a solution of the boundary conditions for each
closed string field $X_S (\sigma )$ with periodicity $2\pi k$.  Multiple
closed strings are constructed via the usual block diagonal procedure.

\eqn{uopen}{
X^i(\sigma^2)=U(\sigma^2)
D
U^{-1}(\sigma^2),}
\eqn{diagstring}{\begin{array}{c}D=\mbox{diag}(X_s^i(\sigma^2),\epsilon
X_s^i(2\pi-\sigma^2),X_s^i(2\pi+\sigma^2),
\epsilon X_s^i(4\pi-\sigma^2),\\
\dots,
X_s^i(2\pi(N-1)+\sigma^2),\epsilon X_s^i(2\pi N-\sigma^2)).
\end{array}}
where $\epsilon$ is $+1$ for $X^{2\dots 9}$ and $\theta$'s,
$-1$ for $A^1$ and $\lambda$'s.
{}From this form it is clear that the matrices will commute
with each other for any value of $\sigma^2$.
We must obey Neumann boundary conditions
for the real part of matrices and Dirichlet conditions
for the imaginary parts (or for $\epsilon=-1$ vice versa),
so we must use specific values of the unitary matrix
$U(\sigma^2)$ at the points $\sigma^2=0,\pi$.
Let us choose
\eq{U'(0+)=U'(\pi-)=0}
(for instance, put $U$ constant on a neighbourhood
of the points $\sigma^2=0,\pi$)
and for a closed string,
\eq{U(\pi)=\tb{{cccc}
m&&&\\
&m&&\\
&&\ddots &\\
&&&m},\qquad
U(0)=C\cdot
U(\pi)\cdot C^{-1},}
where $C$ is a cyclic permutation matrix
\eq{C=\tb{{ccccc}
&1&&&\\
&&1&&\\
&&&\ddots&\\
&&&&1\\
1&&&&}}
where $m$ are $2\times 2$ blocks (there are $N$ of them)
(while in the second matrix the $1$'s are $1\times 1$ matrices so that
we have a shift of the $U(\pi)$ along the diagonal
by half the size of the Pauli matrices.
The form of these blocks guarantees the conversion of $\tau_3$
to $\tau_2$:
\eq{m=\frac{\tau_2+\tau_3}{\sqrt 2}.}
This $2\times 2$ matrix causes two ends to be connected
on the boundary. It is easy to check that the right
boundary conditions will be obeyed.

To obtain open strings, we just change the
$U(0)$ and $U(\pi )$.
An open string of odd length is obtained by
throwing out the last element in (\ref{diagstring}) and taking
\eq{
U(0)=\tb{{ccccc}
1_{1\times 1}&&&&\\
&m&&&\\
&&m&&\\
&&&\dots &\\
&&&&m
},\quad\,
U(\pi)=\tb{{ccccc}
m&&&\\
&m&&\\
&&m&\\
&&&\dots&\\
&&&&1_{1\times 1}}}
Similarly, an open string of even length  will have one of the matrices
$U(0),U(\pi)$ equal to what it was in the closed string case $m\otimes
1$ while
the other will be equal to
\eq{
U(0)={\small\tb{{cccccc}
1_{1\times 1}&&&&&\\
&m&&&&\\
&&m&&&\\
&&&\dots &&\\
&&&&m&\\
&&&&&1_{1\times 1}
}}}

Similar constructions for the fermionic coordinates are straightforward
to obtain.  We also note that we have worked above with the original
boundary conditions and thus obtain only open strings whose ends are
attached to the wall at $R_1 = 0$.  Shifting the boundary condition
\ref{bcc} by $2\pi R_1$ (either at $\sigma^2 =0$ or $\sigma^2 = \pi$ or
both) we obtain strings attached to the other wall, or with one end on
each wall.  Finally, we note that we can perform the gauge
transformation $M \rightarrow \tau_3 M \tau_3$ on our construction.
This has the effect of reversing the orientation of  the string fields,
$X_S (\sigma^2 ) \rightarrow X_S (- \sigma^2 )$.  Thus we obtain
unoriented strings.

We will end this section with a brief comment about moving $D8$ branes
away from the orientifold wall.  This is achieved by adding explicit
$SO(16) \times SO(16)$ Wilson lines to the Lagrangian of the $\chi^a$
fields.  We are working in the regime $R_2 \ll l_{11} \ll R_1$, and we
take these to be constant gauge potentials of the form $\chi^a {\cal
A}_{ab} \chi^b$, with ${\cal A}$ of order $R_1$.  In the presence of
such terms $\chi^a$ will not have any low frequency modes, unless we
also shift the $O(M)$ gauge potential $A_1$ to give a compensating shift
of the $\chi$ frequency.  In this way we can construct open strings
whose ends lie on $D8$ branes which are not sitting on the orientifold.

In this construction, it is natural to imagine that $16$ of the $\chi$
fields live on each of the boundaries of the dual cylinder.  Similarly,
for larger values of $d$ it is natural to put ${32 \over 2^d}$ fermions
on each orbifold circle, a prescription which clearly runs into problems
when $d > 4$. This is reminiscent of other orbifold constructions in M
theory in which the most symmetrical treatment of fixed points is not
possible (but here our orbifold is in the dual world volume).  It is
clear that our understanding of the heterotic matrix model for general
$d$ is as yet quite incomplete.  We hope to return to it in a future
paper.

\section{Conclusions}

We have described a class of matrix field theories which incorporate the
Fock spaces of the the $E_8 \times E_8$
heterotic/Type $IA$ string field theories
into a unified quantum theory.  The underlying gauge dynamics provides a
prescription for string interactions.  It is natural to ask what the
connection is between this nonperturbatively defined system and previous
descriptions of the nonperturbative dynamics of string theories with
twice as much supersymmetry.  Can these be viewed as two classes of
vacua of a single theory?  Can all of these be obtained as different
large $N$ limits of a quantum system with a finite number of degrees of
freedom?

The necessity of introducing the $\chi$ fields into our model suggests
that the original eleven dimensional system does not have all the
necessary ingredients to be the underlying theory.  Yet we are used to
thinking of obtaining lower dimensional compactifications by restricting
the degrees of freedom of a higher dimensional theory in various ways.
Insight into this puzzle can be gained by considering the limit of
heterotic string theory which, according to string duality, is supposed
to reproduce M theory on $K3$.  The latter theory surely reduces
to eleven dimensional M theory in a continuous manner as the radius of
$K3$ is taken to infinity.
Although we have not yet worked out the details of
heterotic matrix theory on higher dimensional tori, we think that it
is clear that the infinite $K3$
limit will be one in which the $\chi$ degrees
decouple from low energy dynamics.

The lesson we learn from this example is that {\it decompactification of
space time dimensions leads to a reduction in degrees of freedom}.
Indeed, this principle is clearly evident in the prescription for
compactification of M theory on tori in terms of SYM theory.  The more
dimensions we compactify, the higher the dimension of the field theory we
need to describe the compactification.  There has been some discussion
of whether this really corresponds to adding degrees of freedom since
the requisite fields arise as limits of finite matrices.  However there
is a way of stating the principle which is independent of how one
chooses to view these constructions. Consider, for example, a graviton
state in M theory compactified on a circle.  Choose a reference energy
$E$ and ask how the number of degrees of freedom with energy less than
$E$ which are necessary to describe this state, changes with the radius
of compactification.  As the radius is increased, the radius of the dual
torus decreases.  This decreases the number of states in the theory with
energy less than $E$, precisely the opposite of what occurs when we
increase the radius of compactification of a local field theory

\subsection{Cosmological Constant Problem}

It seems natural to speculate that this property, so counterintuitive
from the point of view of local field theory, has something to do with
the cosmological constant problem.  In \cite{tbcosmo}
one of the authors
suggested that any theory which satisfied the 't Hooft-Susskind
holographic principle
would suffer a thinning out of degrees
of freedom as the universe expanded, and that this would lead to an
explanation of the cosmological constant problem.  Although the
speculations there did not quite hit the mark, the present ideas suggest
a similar mechanism.  Consider a hypothetical state of the matrix model
corresponding to a universe with some number of Planck size dimensions
and some other dimensions of a much larger size, $R$.  Suppose also that
SUSY is broken at scale $B$, much less than
the (eleven dimensional) Planck scale.
The degrees of freedom
associated with the compactified dimensions all have energies
much higher than the
SUSY breaking scale. Their zero point fluctuations will lead
to a finite, small (relative to the Planck mass) $R$ independent,
contribution to the total vacuum
energy.
As $R$ increases, the number of degrees of freedom at scales less than
or equal to $B$ will decrease.
Thus, we expect a corresponding decrease in the
total vacuum energy.
The total vacuum energy in the large $R$ limit is thus bounded
by a constant, and is dominated by the
contribution of degrees of freedom associated with
the small, compactified dimensions.
Assuming only the minimal supersymmetric cancellation
in the computation of the vacuum energy,
we expect it to be of order $B^2 l_{11}$.  This
implies a vacuum energy density of order
$B^2 l_{11} / R^3$, which is too small to be of
observational interest for any plausible values of the parameters.
If a calculation of this nature turns
out to be correct, it would constitute a prediction
that the cosmological constant is essentially zero in the matrix model.

It should not be necessary to emphasize how premature
it is to indulge in speculations
of this sort (but we couldn't resist the temptation).
We do not understand supersymmetry
breaking in the matrix model and we are even
further from understanding its cosmology.
Indeed, at the moment we do not even
have a matrix model derivation of the fact\footnote{Indeed
this ``fact'' is derived by rather indirect arguments in perturbative
string theory.} that
parameters like the radius of compactification are dynamical variables.
Perhaps the
most important lacuna in our understanding
is related to the nature of the large $N$
limit.  We know that many states of the system
wander off to infinite energy as $N$
is increased.
Our discussion above was based on extrapolating results of the finite $N$
models, without carefully verifying that the
degrees of freedom involved survive
the limit.  Another disturbing thing about our discussion is the absence
of a connection to Bekenstein's area law for the number of states.  The
Bekenstein law
seems to be an integral part of the physical picture of the matrix model.
Despite these obvious problems, we feel
that it was worthwhile to present this preliminary discussion of
the cosmological constant problem
because it makes clear that the spacetime picture
which will eventually emerge from the matrix model is certain to be
very different from the one implicit in local field theory.

\section{Acknowledgements}
L.Motl is grateful to staff and students of Rutgers University
for their hospitality.

\newpage


\begin{thebibliography}{19}        

\bibitem[1]{dhn}B.\,de Wit, J.\, Hoppe, H.\,Nicolai,
{\it Nucl. Phys.}{\bf B305} [FS 23], (1988), 545.
\bibitem[2]{towns} P.K.\,Townsend
{\it Phys. Lett.}{\bf B373}, (1996), 68, hep-th/9512062.
\bibitem[3]{bfss}T.\,Banks, W.\,Fischler,\,S.H.\,Shenker,
L.\,Susskind, hep-th/9610043
\bibitem[4]{orbi} U.\,Danielsson, G.\,Ferretti, hep-th/9610082;\newline
S.\,Kachru, E.\,Silverstein, hep-th/9612162;\newline
L.\,Motl, hep-th/9612198;\newline
D.\,Lowe, hep-th/9702006;\newline
N.\,Kim, S.J.\,Rey, hep-th/9701139;\newline
T.\,Banks, N.\,Seiberg, E.\,Silverstein, hep-th/9703052.
\bibitem[5]{evashamit}S.\,Kachru, E.\,Silverstein, hep-th/9612162
\bibitem[6]{df}U.\,Danielsson, G.\,Ferretti, hep-th/9610082
\bibitem[7]{motl}L.\,Motl, hep-th/9612198
\bibitem[8]{jopo}J.\,Polchinski, E.\,Witten,
{\it Nucl. Phys.}{\bf B460},(1996), 525, hep-th/9510169.
\bibitem[9]{rk}N.\,Kim, S.J.\,Rey, hep-th/9701139
\bibitem[10]{bss}T.\,Banks, N.\,Seiberg,
E.\,Silverstein, hep-th/9703052.
\bibitem[11]{bs}T.\,Banks, N.\,Seiberg,
hep-th/9702187
\bibitem[12]{BPS}J.\,Maldacena, L.\,Susskind,
{\it Nucl. Phys.} {\bf B475}, (1996), 679;\newline
S.\,Das, S.\,Mathur,
{\it Nucl. Phys.}{\bf B478}, (1996), 561, hep-th/9606185;\newline
G.\,Moore, R.\,Dijkgraaf, E.\,Verlinde, H.\,Verlinde, hep-th/9608096.
\bibitem[13]{tbcosmo}T.\,Banks, hep-th/9601151.
\bibitem[14]{taylor}W.\,Taylor, hep-th/9611042
\bibitem[15]{motlb}L.\,Motl, hep-th/9701025
\bibitem[16]{dvv}R.\,Dijkgraaf, E.\,Verlinde, H.\,Verlinde, hep-th/9703030



\end{thebibliography}
\end{document}